\documentstyle{l-aa}
\input{psfig.txt}

\def\kms{\,km\,s$^{-1}$}

\def\lsol{~L$_\odot$ }

\def\micron{\,$\mu$m}
 
\def\marc{mag~arcsec$^{-2}$}

\begin{document}

\thesaurus{11.06.2;
	   11.16.1;
	   13.09.1
	   }

\title{1.65~$\rm \mu$m (H-band) surface photometry of galaxies. V: 
Profile decomposition of 1157 galaxies.\thanks{Based on observations 
taken at TIRGO, Gornergrat, Switzerland (operated by CAISMI-CNR,
Arcetri, Firenze, Italy) and at the Calar Alto Observatory (operated
by the Max-Planck-Institut f\"ur Astronomie (Heidelberg) jointly with
the Spanish National Commission for Astronomy).}
}


\author{G. Gavazzi\inst{1}
\and P. Franzetti \inst{1}
\and M. Scodeggio \inst{2}
\and A. Boselli\inst{3}
\and D. Pierini\inst{4}  
}

\offprints{G. Gavazzi}

\institute{Universit\`a degli Studi di Milano - Bicocca, P.zza 
dell'Ateneo Nuovo 1, 20126 Milano, Italy
\and 
Istituto di Fisica Cosmica ``G. Occhialini'', CNR, via Bassini 15, 
20133, Milano, Italy
\and
Laboratoire d'Astronomie Spatiale, Traverse du Siphon, F-13376 
Marseille Cedex 12, France
\and
Max Planck Institut f\"ur Kernphysik, postfach 103980, D-69117 
Heidelberg, Germany
}

\date{Received..........; accepted..........}

\maketitle

\markboth{G. Gavazzi et al.: NIR surface brightness profiles of 1157 
galaxies}{}

\begin{abstract}

We present near-infrared H-band (1.65\micron ) surface brightness
profile decomposition for 1157 galaxies in five nearby clusters of
galaxies: Coma, A1367, Virgo, A262 and Cancer, and in the bridge
between Coma and A1367 in the "Great Wall".  The optically selected
($m_{pg}\leq 16.0$) sample is representative of all Hubble types, from E
to Irr+BCD, except dE and of significantly different
environments, spanning from isolated regions to rich clusters
of galaxies.  We model the surface brightness profiles with
a de Vaucouleurs $r^{1/4}$ law (dV), with an exponential disk law (E),
or with a combination of the two (B+D).  From the fitted quantities we
derive the H band effective surface brightness ($\mu_e$) and radius
($r_e$) of each component, the asymptotic magnitude $H_T$ and the
light concentration index $C_{31}$.  We find that: i) Less than 50\%
of the Elliptical galaxies have pure dV profiles.  The majority of E
to Sb galaxies is best represented by a B+D profile.  All Scd to BCD
galaxies have pure exponential profiles.  ii) The type of
decomposition is a strong function of the total H band luminosity
(mass), independent of the Hubble classification: the fraction of pure
exponential decompositions decreases with increasing luminosity, that
of B+D increases with luminosity.  Pure dV profiles are absent in the
low luminosity range $L_H<10^{10}$ \lsol and become dominant above
$10^{11}$ \lsol.

\end{abstract}

\section{Introduction}

Since the advent of large format near-infrared (NIR) arrays, extensive
surface photometry of galaxies has been carried out in the NIR
domain. De Jong \& van der Kruit (1994) did observations of 86 spiral
galaxies and similar observations were obtained by Block et
al. (1994).  Since the NIR is the most suitable band for studying the
properties which depend of the old stellar population in galaxies,
unaffected by recent episods of star formation, many observational
studies were devoted to early-type galaxies, with the aim of studying
their NIR fundamental plane (e.g. Pahre, 1999).  Very little work
exists in the literature at NIR passbands addressing at the same time
the properties of early and late-type galaxies.  To fill this gap,
since 1993, we made extensive use of NIR panoramic detectors to obtain
H (and K') band images of galaxies.  We first concentrated on disk
galaxies (see Gavazzi et al. 1996a (Paper I), Gavazzi et al. 1996b
(Paper II), Boselli et al. 2000 (Paper IV) and Boselli et al. 1997
(B97)), later we extended the survey to the early-types (Gavazzi et
al. 2000 (Paper III)).  The observing sample was selected among
members of 5 nearby, rich clusters: namely the Virgo, Coma, A1367,
A262 and Cancer clusters.  In addition, a significant population of
galaxies in the ``Great Wall'', the bridge between Coma and A1367, was
included.  The present survey, due to its completeness, can be
considered representative of the NIR properties of nearby galaxies,
both of early and late-types.  Moreover the complete coverage of 5
clusters and of all members of the "Great Wall" (see Gavazzi et
al. 1999) makes it possible to study the environmental dependence of
the NIR properties of galaxies.  We reiterate, however, that the
present survey is not composed of NIR selected galaxies, rather it
contains NIR observations of optically selected objects.  In this
paper we concentrate on the structural properties of galaxies that can
be derived from surface-photometry measurements: i.e.  on the light
profiles of galaxies at NIR pass-bands.  The paper is organized as
follows: the sample selection criteria are discussed in Section 2 and
the procedures adopted to derive the light profiles and the models
fitted to the data are given in Section 3.  The results of the present
work, their internal and external consistency are given in Section
4. Some implications of the results of the present analysis on the
structural properties of galaxies are discussed in Section 5 and
summarized in Section 6.

\section {Sample selection}

\begin{table*}
\caption{Sample completeness}
\label{Tab1}
\[
\begin{array}{p{0.35\linewidth}cccccc}
\hline
\noalign{\smallskip}
Region   	            &  E+L &      & Early&     & Late &     \\
                            &  N   & (\%)   & N   &  (\%)  & N   &  (\%)  \\
 (1)                        & (2)  &  (3) & (4) &  (5) & (6) & (7)  \\
\noalign{\smallskip}
\hline
\noalign{\smallskip}
ComaSup "Bridge"(Isol)      &  374 & (96)   & 194 & (97)   & 180 & (95)   \\
A1367+A1656                 &  251 & (97)   & 179 & (97)   &  72 & (99)   \\
A262                        &   68 & (67)   &  10 & (24)   &  58 & (95)   \\
Cancer                      &   57 &(100)   &  20 &(100)   &  37 &(100)   \\
Virgo $m_p<14.0$            &  220 & (89)   &  81 & (83)   & 139 & (93)   \\
Virgo $m_p<16.0$            &  276 & (47)   &  86 & (30)   & 190 & (62)   \\
Virgo $14.0<m_p<16.0$ (ISO) &      &        &     &        &  38 & (97)   \\
Miscellanea                 &  131 &        &     &        &     &        \\        
\noalign{\smallskip}
\hline
\end{array}
\]
\end{table*}

The NIR observations analyzed in this paper are taken from Paper I,
II, III, IV of this series and from B97.  They were obtained from 1993
to 1997 with the 1.5~m TIRGO and with the Calar Alto 2.2 and 3.5~m
telescopes equipped with the NICMOS3 $256^2$ pixel arrays cameras
ARNICA (Lisi et al. 1993) and MAGIC (Herbst et al. 1993),
respectively.  In total we have analyzed 1285 images. However, by
selecting the best quality images among 128 repeated measurements, the
total number of independent observations of individual galaxies
reduces to 1157. \\ 
The observed galaxies were optically selected from either the CGCG
catalogue (Zwicky et al. 1961-68) (m$\rm _p \le$~15.7) or from the VCC
catalogue (Binggeli et al. 1985) (restricted to m$\rm _p \le$~16.0).
They belong to the Coma supercluster region: 
($\rm 18^o \le \delta \le 32^o$; $\rm 11.5^h \le \alpha \le 13.5^h$), 
to the A262: 
($\rm 34.5^o \le \delta \le 38.5^o$; $\rm 01^h43^m \le \alpha \le 02^h01^m$),
Cancer: 
($\rm 20.5^o \le \delta \le 23.0^o$; $\rm 08^h11^m \le \alpha \le 08^h25^m$) 
and Virgo cluster: 
($\rm 0.0^o \le \delta \le 20.0^o$; $\rm 12^h00^m \le \alpha \le 13^h00^m$).\\ 
Out of the 1157 galaxies observed in these regions, 1026 constitute a
basically complete sample (see below).  The remaining 131 observations
belong to an incomplete set of data. \\ 
The complete sub-set is composed as follows (see Table 1 for more
details): out of the 646 galaxies, of both early and late-types that
are members to the Coma supercluster according to Gavazzi et
al. (1999), i.e. 5000 $<$ V $<$ 8000 \kms, 625 (97 \%) have been
imaged.  
Among the Coma supercluster objects, 374 (60\%) are galaxies belonging to
the "bridge" between Coma and A1367. These can be treated as 
isolated objects because they inhabit regions of local density
$\sim$ 10 times lower than rich clusters (see Gavazzi et
al. 1999). Their completeness allows us to use them as
a reference sample for studying the environmental dependence
of the galaxy structural properties, in comparison with those
of the rich cluster sample.\\
A 100\% complete imaging coverage exists as well for galaxies in the
Cancer cluster, while the late-type galaxies in the A262 cluster were
covered in a quasi-complete manner (95\%), as opposed to the
early-types ones (24\% complete).  Moreover the survey contains 220
out of 248 (89 \% complete) VCC galaxies brighter than $m_p$=14.0.
Thus the giant members of the Virgo cluster (V $<$ 3000~ \kms) are
sampled in a quasi-complete manner.  A less complete coverage exists
at $m_p~\leq~16.0$: 276/587 objects (47 \% complete).  However, we
have observed all but one the 88 late-type VCC galaxies selected by
the ISO consortium (B97) brighter than $m_p$=16.0. These are objects
lying either within 2 degrees of projected radial distance from M87 or
in the corona between 4 and 6 degrees (remark that the observations in
B97 were carried on in the K' band).  Thus the NIR survey contains a
representative sample of late-type dwarf ($m_p~\leq~16.0$) members of
the Virgo cluster, restricted however to a region smaller than the
whole VCC. \newline 
The incomplete sample of 131 objects mainly comprises galaxies
projected onto the Coma supercluster region, either on the background
(V $>$ 8000 \kms), or in the foreground (V $<$ 5000 \kms) (121
objects), or in the background of the Virgo cluster (V $>$ 3000
\kms)(10 objects).\\ 
Unfortunately the survey does not presently cover dwarf-elliptical
galaxies.  Only 6 such objects were serendipitously observed so far.

\section{Data reduction procedures}

Papers I, II, III, IV of this series and B97 give all the details of
the observations and the methods used in preparing the flat-fielded,
combined, and calibrated frames used in the present analysis, which
was performed in the IRAF environment and relied on the STSDAS package
\footnote{IRAF is the Image Analysis and Reduction Facility made
available to the astronomical community by the National Optical
Astronomy Observatories, which are operated by AURA, Inc., under
contract with the U.S. National Science Foundation. STSDAS is
distributed by the Space Telescope Science Institute, which is
operated by the Association of Universities for Research in Astronomy
(AURA), Inc., under NASA contract NAS 5--26555.}  and on GALPHOT
(developed for IRAF- STSDAS mainly by W. Freudling, J. Salzer, and
M.P. Haynes and adapted by us to handle NIR data).\\ 
For each frame the sky background was determined as the mean number of
counts measured in regions of ``empty'' sky, and it was subtracted
from the frame.  Sky-subtracted frames were inspected individually and
the light of unwanted superposed or nearby stars and galaxies was
masked.\\ 
The 2-dimensional light distribution of each galaxy was fitted with
elliptical isophotes, using a modified version of the STSDAS ${\it
isophote}$ package.  Starting from a set of initial parameters given
manually, the fit maintains as free parameters the ellipse center,
ellipticity and position angle. The ellipse semi-major axis is
incremented by a fixed fraction of its value at each step of the
fitting procedure.  The routine halts when the surface brightness
found in a given corona equals the sky rms.  The fit fails to converge
for some galaxies with very irregular light distributions.  In these 
cases we keep fixed one or more of the initial parameters. \\ The
resulting radial light profiles are fitted with models of the galaxy
light distribution: a de Vaucouleurs $r^{1/4}$ law (de Vaucouleurs
1948) or an exponential disk law, or a combination of the two.\\

\begin{figure*}
\psfig{figure=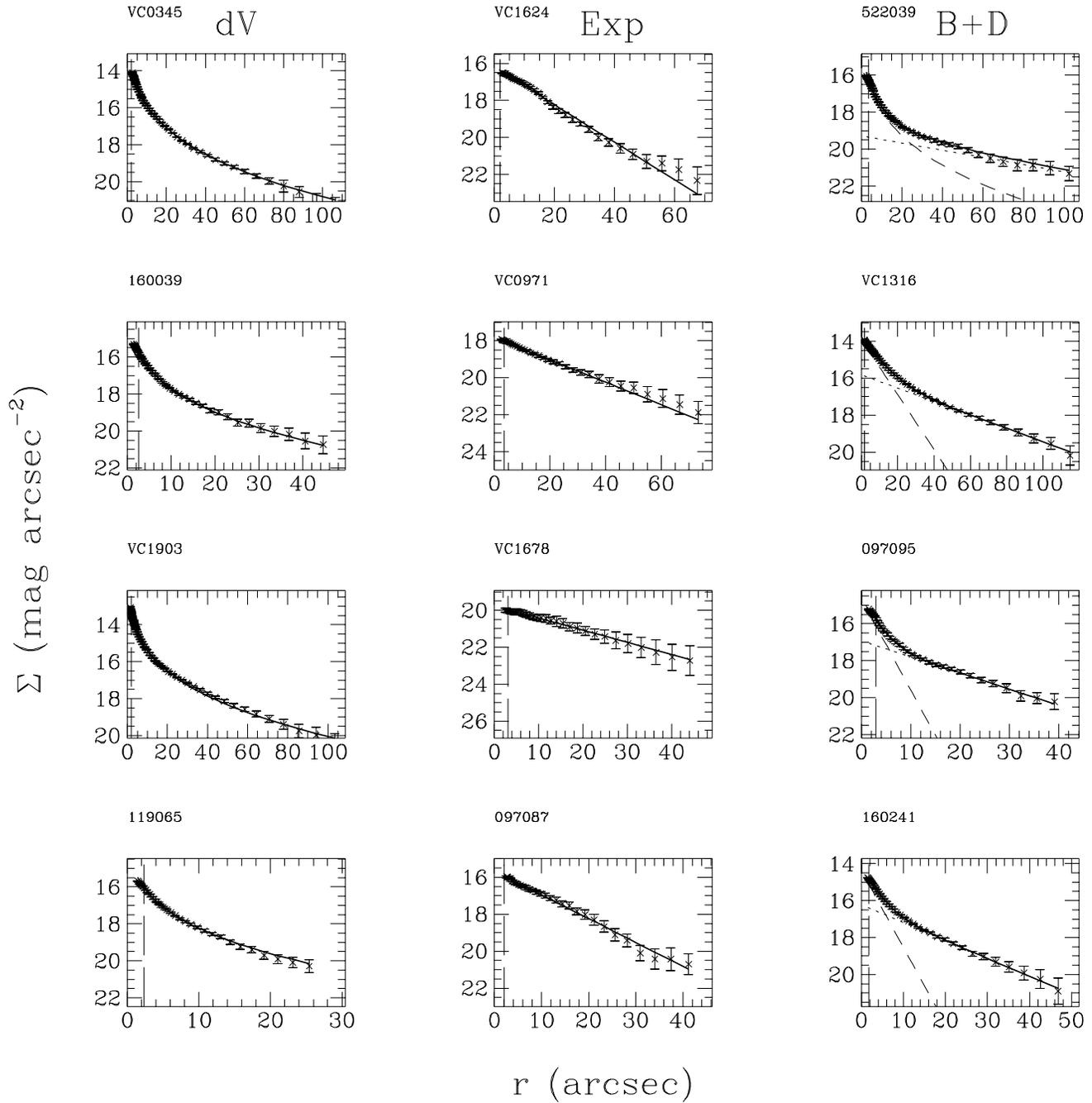,width=19cm,height=19cm}
\caption{Examples of profile decompositions of 12 well known galaxies.
Four galaxies with a pure de Vaucouleurs decomposition are shown in
the left column: These are (from top to bottom): VCC345=N4261 (E in
Virgo), 160039=N4839 (E in Coma), VCC1903=N4621 (E in Virgo) and
119065=N2563 (the brighest E in Cancer).  Four galaxies with a pure
exponential decomposition are shown in the middle column: These are:
VCC1624=N4544 (Sc in Virgo), VCC971=N4423 (Sd in Virgo),
VCC1678=IC3576 (Sd in Virgo) and 97087=U6697 (the brightest Irr/Pec in
A1367).  Four galaxies with a mixed Bulge+Disk decomposition are shown
in the right column: These are: 522039=N708 (the brightest E in A262),
VCC1316=N4486=M87, 97095=N3842 (the brightest E in A1367) and
160241=N4889 (the brightest E in Coma).  A vertical broken line is
drawn at the radius of the seeing.  }
\label{fig.1}
\end{figure*}

Among the 1157 profiles, 165 are fitted with a pure de Vaucouleurs
$r^{1/4}$ law and 322 with a pure exponential. The remaining require a
Bulge+Disk (B+D) decomposition. For these we developed an algorithm to
separate the disk from the bulge component, following Kormendy (1977).
We begin by fitting the outer part of the profile (where the disk
component dominates) with an exponential law; then we extrapolate this
fit to the inner region of the galaxy and we subtract it from the
data.  The resulting surface brightness profile is fitted again with
an exponential or with a de Vaucouleurs law, according to a $\chi^2$
test.  This procedure is iterated until the sum of the two components
gives the minimum $\chi^2$.  The fits are performed from a radius
equal to twice the seeing disk, out to the outermost significant
isophotes.\\ 
Total magnitudes $H_T$ are then obtained by adding to the flux
measured within the outermost significant isophote the flux
extrapolated to infinity along either the $r^{1/4}$ (dV galaxies), or
the exponential law that fitted the outer parts of the galaxy (pure
disks and B+D galaxies).  The median uncertainty in the determination
of the total magnitude is 0.15 mag.\\ 
The effective radius $r_e$ (the radius containing half of the total
light) and the effective surface brightness $\mu_e$ (the mean surface
brightness within $r_e$) of each galaxy are computed in two ways: 1)
the "fitted" values ($r_{edf}$, $r_{ebf}$, $\mu_{edf}$, $\mu_{ebf}$),
are derived from the individual fitted profiles, extrapolated to zero
and to infinity. In case of a B+D galaxy, we compute also $r_{ef}$
and $\mu_{ef}$ by integrating to infinity the flux along the two 
sub-profiles independently, adding the two contributions and computing 
the radius at half the total light and the mean surface brightness within 
that radius; 2) the
"empirical" values ($r_e$, $\mu_e$) are obtained locating the half
light point along the observed light profile, where the total amount
of light is given by the total magnitude $H_T$ described above. 
The two determinations are compared in Section 4.4.  The median
uncertainty on the determination of log~$r_e$ and $\mu_e$ is 0.05 and
0.16 mag, respectively.\\ 
Finally we compute other useful parameters: the concentration index
($C_{31}$), defined in de Vaucouleurs (1977) as the model--independent
ratio between the radii that enclose 75\% and 25\% of the total light
$H_T$; the bulge flux to total flux ratio (B/T) and the NIR isophotal
radius $r_H (20.5)$ determined in the elliptical
azimuthally--integrated profiles as the radius at which the surface 
brightness reaches 20.5 H--\marc.\\ 
Some (37) galaxies show ``truncated'' profiles in the outer regions,
i.e. the slope of their profiles increases outward.  For these objects
we fit only the outer part of the profile with an exponential law, to
allow an estimate of their total magnitude, effective radius,
effective surface brightness and $C_{31}$.

\section{Results}

\begin{figure*}
\psfig{figure=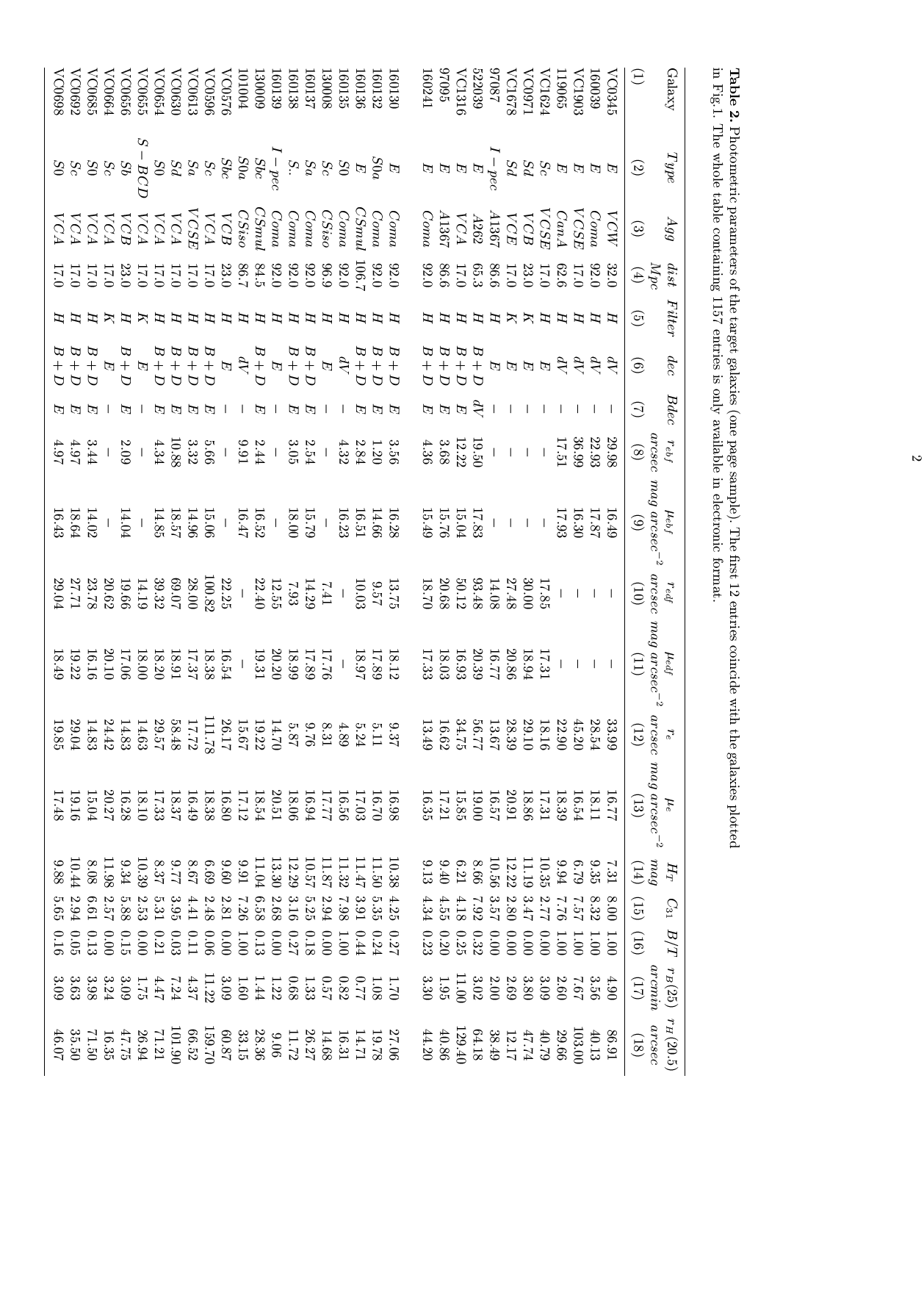,width=18cm,height=23cm}
\end{figure*}

\begin{figure*}
\psfig{figure=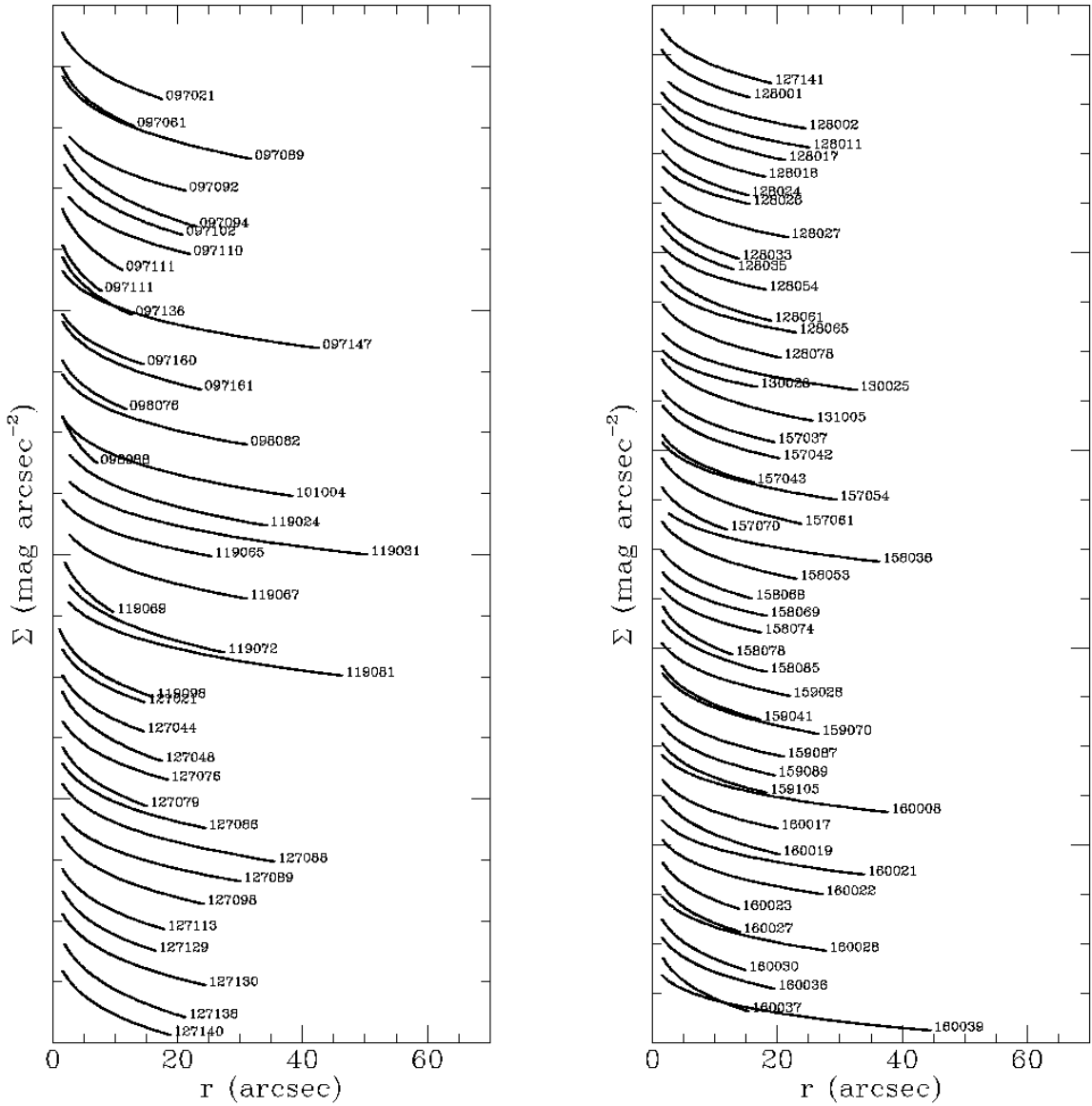,width=18cm,height=18cm}
\caption{A sample of the fitted surface brightness profiles. This
figure illustrates pure de Vaucouleurs profiles.}
\label{fig.2}
\end{figure*}

\begin{figure*}
\psfig{figure=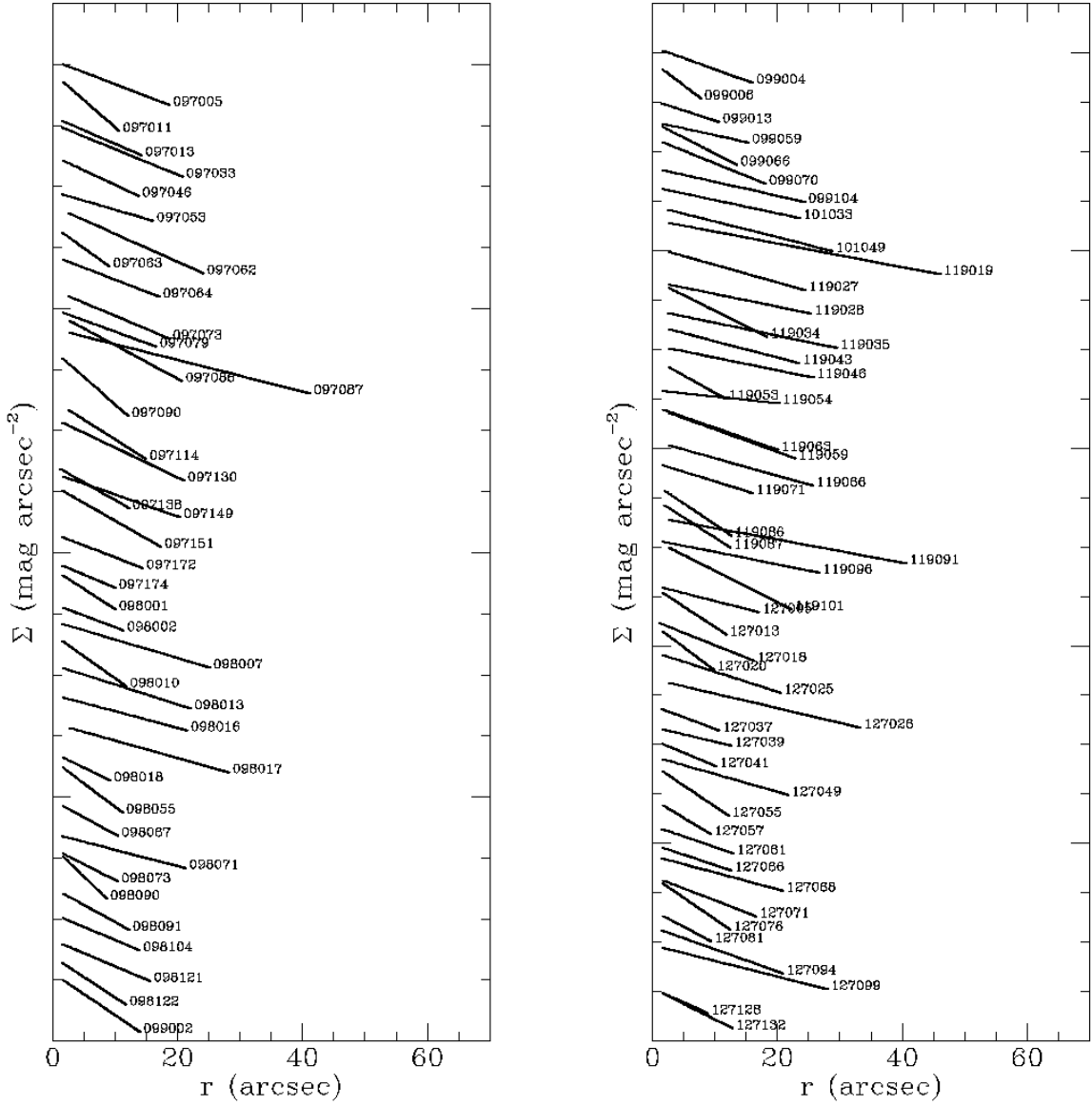,width=18cm,height=18cm}
\caption{A sample of the fitted surface brightness profiles. This
figure illustrates pure exponential profiles.}
\label{fig.3}
\end{figure*}

\begin{figure*}
\psfig{figure=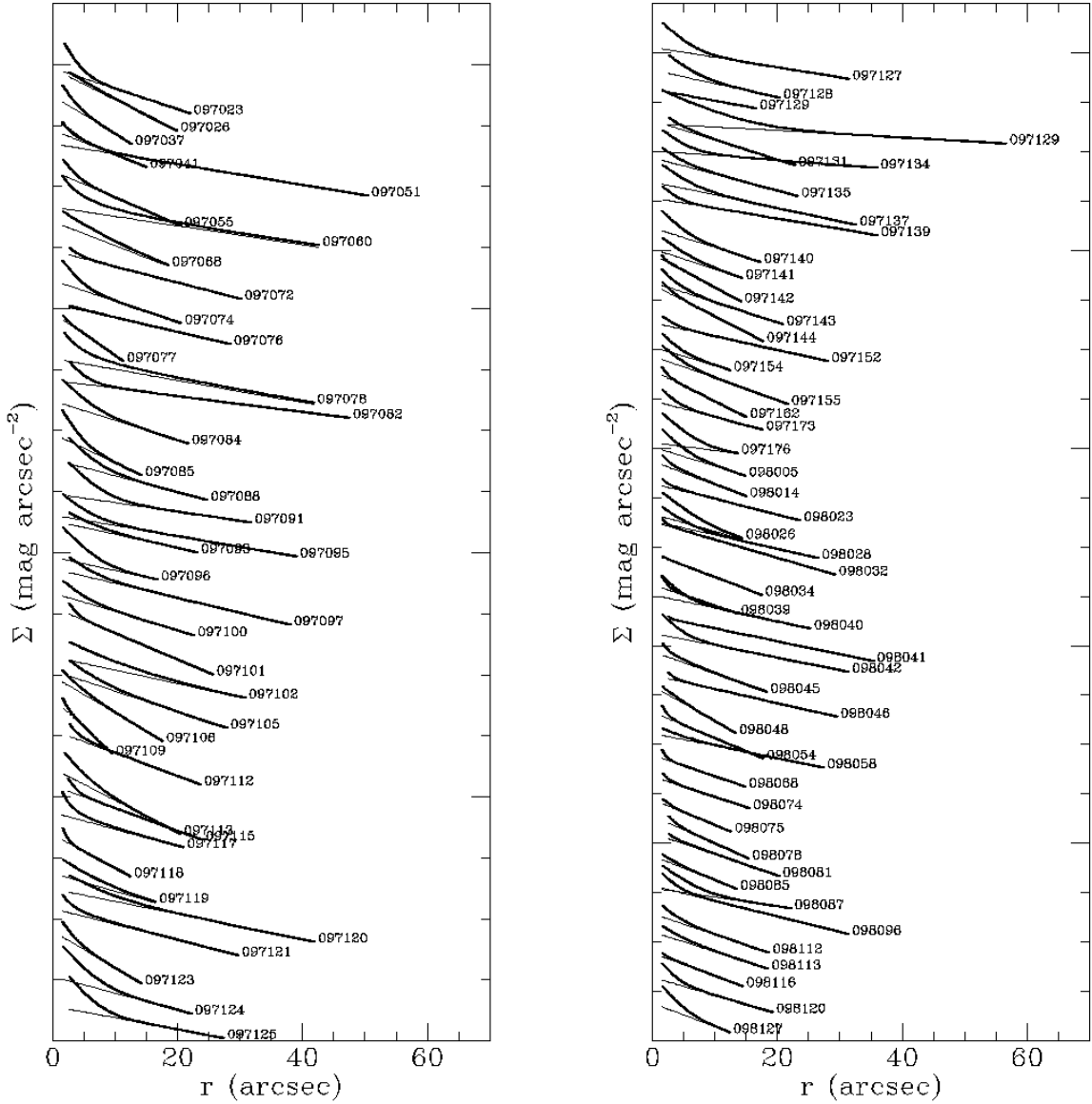,width=18cm,height=18cm}
\caption{A sample of the fitted surface brightness profiles. This
figure illustrates composite profiles.}
\label{fig.4}
\end{figure*}

The results of the present work are summarized in Table
2\footnote{Table 2 and Figs. 2, 3 , 4 are available in their entirety only in electronic
form at the CDS via anonymous ftp to cdsarc.u-strasbg.fr
(130.79.128.5) or via http://cdsweb.u-strasbg.fr/Abstract.html} (here
only one sample page is presented, the whole table is available only
in digital format) as follows: \newline
Column  1: CGCG (Zwicky et al. 1961-68) or VCC (Binggeli et al. 1985) 
denomination.  \newline
Column  2: morphological type. \newline 
Column 3: ``aggregation'' parameter. This parameter defines the
membership to a group/cluster/supercluster: CSisol, CSpairs, CSgroups
indicate members of the Coma Supercluster (5000 $<$ V $<$ 8000 \kms);
CSforeg means objects in the foreground of the Coma Supercluster (V
$<$ 5000 \kms) and CSbackg means objects in the background of the Coma
Supercluster (V $>$ 8000 \kms). Galaxies in the Virgo region are
labeled following the membership criteria given by Binggeli et
al. (1993): VCA, VCB, VCM, VCW, VCSE, VCmem, are members to the
cluster A or B, to the M, W or South-East clouds or are not better
specified members to the Virgo cluster respectively. noVCC are
galaxies taken from the CGCG in the outskirts of Virgo, but outside
the area covered by the VCC.  Subgroups in the Cancer cluster are
identified according to Bothun et al. (1983). \newline 
Column 4: adopted distance in Mpc, using $H_o=75~kms^{-1}Mpc^{-1}$\newline
Column 5: adopted filter (H or K'). \newline 
Column 6: type of decomposition: dV = pure de Vaucouleurs; E = pure
exponential; B+D = Bulge+Disk; T = truncated.\newline 
Column 7: type of decomposition of the bulge: D = de Vaucouleurs; E =
exponential.\newline 
Column 8: effective radius of the fitted bulge component ($r_{ebf}$)
in arcsec.\newline 
Column 9: effective surface brightness of the fitted bulge component
($\mu_{ebf}$) in \marc.\newline 
Column 10: effective radius of the fitted disk component ($r_{edf}$)
in arcsec.\newline 
Column 11: effective surface brightness of the fitted disk component
($\mu_{edf}$) in \marc.\newline 
Column 12: total effective radius ($r_e$) in arcsec (corrected for
seeing according to Saglia 1993).\newline 
Column 13: total effective surface brightness ($\mu_e$) (corrected for
seeing according to Saglia, 1993) in \marc.\newline 
Column 14: total H magnitude ($H_T$) extrapolated to infinity.\newline 
Column 15: concentration index ($\rm C_{31}$) (corrected for seeing
according to Saglia, 1993).\newline 
Column 16: bulge to total flux ratio (B/T).\newline 
Column 17: for CGCG galaxies this is the major optical diameter (r$_B
(25)$) (in arcmin) derived as explained in Gavazzi \& Boselli (1996).
These diameters are consistent with those given in the RC3.  For VCC
galaxies this is the diameter measured on the du Pont plates at the
faintest detectable isophote, as listed in the VCC. \newline 
Column 18: galaxy observed major ($r_H (20.5)$) radius (in arcsec) at
the 20.5 H--\marc~ isophote. Galaxies which require an extrapolation
larger than 0.5 mag to reach the $\rm 20.5^{th}$ magnitude isophote
are labeled -1.\newline 

Fig. 1 gives, as an example, the profile decompositions of 12 well
known galaxies. Figs. 2-4 report, in a more compact form, a sample of
the decompositions obtained in the present work. First galaxies with
pure de Vaucouleurs profiles are presented, then galaxies with pure
exponential disk profile, and finally galaxies that require a B+D
decomposition of their profile. Within each class, the profiles are
ordered with increasing designation number, and are adjusted one after
the other by scaling by one mag their central surface brightness.

\subsection{Consistency of multiple measurements}

Repeated measurements are available for 128 galaxies. From these we
can check the consistency of our procedure.  The quantities:
$\Delta(H_T)$ (difference in asymptotic magnitude), $\Delta(\mu_e)$
(difference in effective surface brightness), $r_{e1}/r_{e2}$
(effective radii ratios) are plotted in Fig. 5 as a function of the
asymptotic magnitude.  The average ratio of radii is 
$<r_{e1}/r_{e2}>=0.97\pm0.17$; the average difference in surface
brightness is $<\Delta(\mu_e)>=-0.08\pm0.30$, while that
in the total magnitude is $<\Delta(H_T)>=-0.01\pm0.22$,
in agreement with the quoted rms = 0.15 mag.

\begin{figure}
\psfig{figure=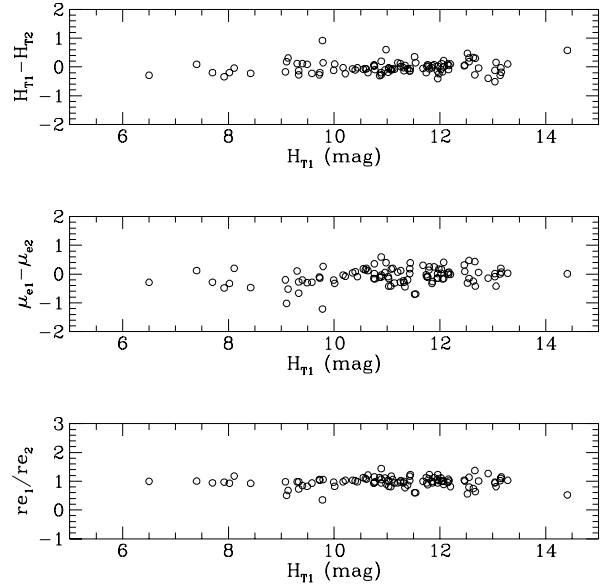,width=12cm,height=12cm}
\caption{Consistency among galaxies with repeated measurement.}
\label{fig.5}
\end{figure}

\subsection{Comparison with independent measurements}

We can compare our decompositions with those obtained in K' band by
Pahre (1999), using 72 early-type galaxies in common, of which 54
belong to the Coma cluster and 18 to the Virgo cluster (see Fig. 6).
To do so we transform the K' magnitudes given by Pahre (1999) into H
magnitudes, using either measured H-K' colours, if available, or
$<H-K'>$=0.25 mag. It is well known however that external comparisons
like this one are affected significantly by systematic uncertainties
introduced by the computational method employed to derive effective
radii and effective surface brightness (see, for example, Scodeggio,
Giovanelli \& Haynes 1998). In fact, we find a rather large scatter in
the galaxy-to-galaxy comparison, and significant differences in the
determination of effective radii. The empirical effective radii
determined in the present work are on average 21 \% larger than Pahre
(1999) ($<r_e/r_{ep}>=1.21\pm0.53$), while the determination of
effective surface brightness and total magnitude are in much better
agreement ($<\Delta(\mu_e)>=+0.04\pm0.84$; $<\Delta(H_T)>=0.04\pm0.25$).  
However if we compare our $H_B (25)$ magnitudes (i.e. the H band
magnitudes extrapolated to the optical radius, as given in the data
papers of this series) we find a somewhat reduced scatter of 0.21 mag
(in agreement with an rms = 0.15 mag on each party).  This is not
surprising since the extrapolation along the model profile carries its
own contribution to the total photometric error.
From Fig. 6 it is evident that the largest discrepancies occur for
the faintest, and therefore smallest, galaxies. This might indicate
the presence of some residual seeing effect in either one of the two
datasets. 

\begin{figure}
\psfig{figure=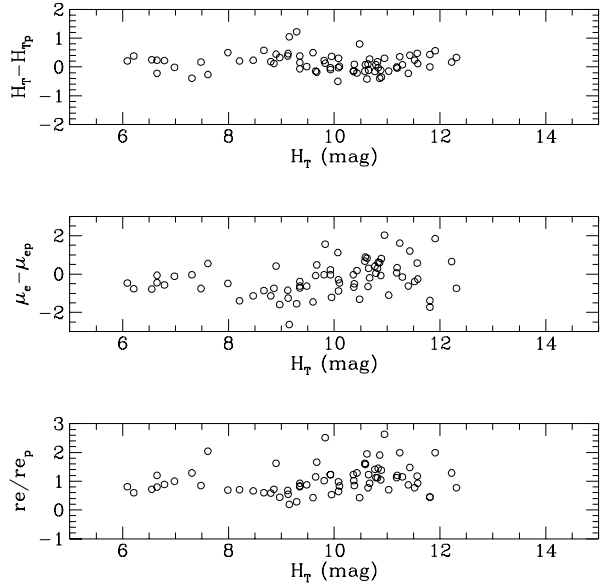,width=12cm,height=12cm}
\caption{Consistency with Pahre (1999)}
\label{fig.6}
\end{figure}

\subsection{Effects of the seeing}

The seeing produces significant light smearing in astronomical images,
which systematically makes their appearance less centrally
peaked. This increasingly biases the statistical distribution of the
light profiles against pure de Vaucouleurs decompositions of galaxies
of decreasing size.  Since the seeing in the present observations, in
particular the ones carried out at TIRGO, was far from optimal (see
e.g. paper III) we have considered this effect carefully.  To this aim
we take the following steps: 1) the fitting algorithms are run outside
twice the seeing disk.  2) empirical $r_e$, $\mu_e$ and $C_{31}$ of
galaxies with relevant central cusps (pure de Vaucouleurs and with
$B/T>0.5$) are corrected for seeing following the prescriptions of
Saglia (1993).  We have checked that the corrected quantities do not
contain any residual dependence on the seeing.

\subsection{Consistency of empirical vs. fitted quantities}

The "empirical" parameters $\mu_e$ and $r_e$ are compared in Fig. 7
and 8 with the corresponding fitted values for the three classes of
decompositions (i.e. $r_{edf}$ and $\mu_{edf}$ for pure exponential fits,
$\mu_{ebf}$ and $r_{ebf}$ for pure de Vaucouleurs, $\mu_{ef}$ and $r_{ef}$ 
for B+D galaxies. These quantities are indicated as $\mu_{f}$ 
and $r_{f}$ for simplicity in Figs. 7 and 8).\\
For pure exponential profiles there is excellent agreement between the
measured quantities (see Table 2). A satisfactory agreement exists
also for B+D profiles.  The pure de Vaucouleurs profiles present a
systematic difference: the empirical surface brightness are half a
magnitude fainter and the corresponding radii 23\% larger than the
fitted quantities.  This is not an unexpected result, instead it
derives from our fitting strategy: in fact, in order to avoid the
effects of the seeing, we choose to mask the data in the inner regions
(up to a radius equal to twice the seeing disk) during the fitting
procedure.  Consequently the fit exceeds the measured surface
brightness in the central parts, meanwhile the fitted $r_f$ turns out
smaller than $r_e$.\\ 
For galaxies with exponential or mixed profiles the difference between
the total magnitude extrapolated along the fit and that truncated at
the optical $r_B (25)$ radius is only 0.1 mag on average, as expected.
For de Vaucouleurs profiles this average discrepancy becomes 0.30 mag,
because de Vaucouleurs profiles contain a significant flux
contribution from the outer parts (see Fig. 9).  Whether Elliptical
galaxies follow de Vaucouleurs profiles up to large radii or milder
exponential "truncations" exist in the outer parts is a debated issue.
Certainly a significant fraction of Elliptical galaxies in the present
work show exponential (H band) outer profiles (see section 5.2).\\
\begin{figure}
\psfig{figure=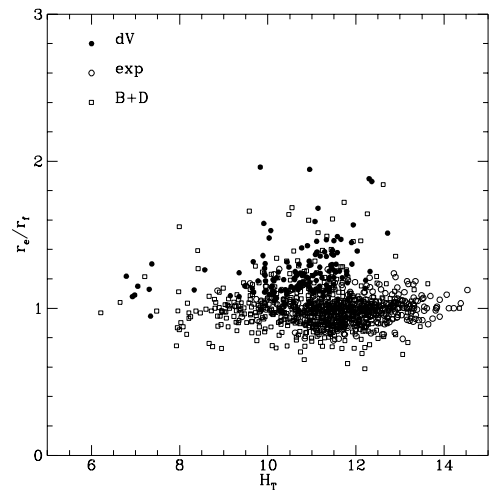,width=10cm,height=10cm}
\caption{The ratio $r_e/r_f$ as a function of $H_T$.}
\label{fig.7}
\end{figure}

\setcounter{table}{3}
\begin{table*}
\caption{comparison of empirical vs. fitted quantities}
\label{TaCapab2}
\[
\begin{array}{p{0.25\linewidth}ccc}
\hline
\noalign{\smallskip}
Decomp    & \mu_e-\mu_f   & r_e/r_f     & H_B(25)-H_T  \\
 (1)        & (2)         &    (3)      &      (4)     \\
\noalign{\smallskip}
\hline
\noalign{\smallskip}
All (1102)  & 0.08\pm0.34 & 1.03\pm0.16 & 0.09\pm0.20  \\
dV  (161)   & 0.45\pm0.33 & 1.23\pm0.18 & 0.30\pm0.19  \\
Exp (322)   & 0.00\pm0.18 & 1.00\pm0.09 & 0.04\pm0.19  \\
B+D (619)   & 0.01\pm0.34 & 0.99\pm0.15 & 0.06\pm0.17  \\
\noalign{\smallskip}
\hline
\end{array}
\]
\end{table*}

\begin{figure}
\psfig{figure=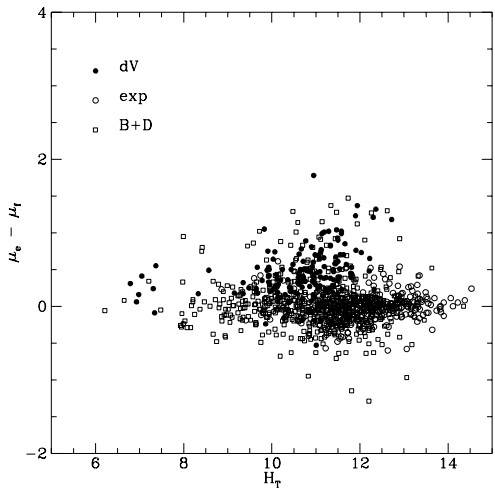,width=10cm,height=10cm}
\caption{The difference $\mu_e-\mu_f$ as a function of  $H_T$.}
\label{fig.8}
\end{figure}

\begin{figure}
\psfig{figure=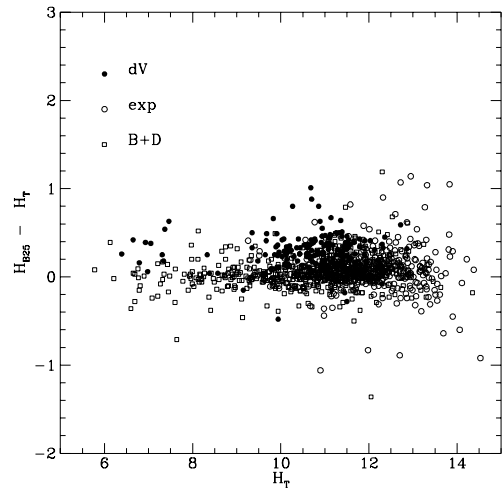,width=10cm,height=10cm}
\caption{The difference $H_B(25)-H_T$ as a function of $H_T$.}
\label{fig.9}
\end{figure}

\section{Analysis}

\subsection{Isophotal, empirical, and optical radii}

\begin{figure}
\psfig{figure=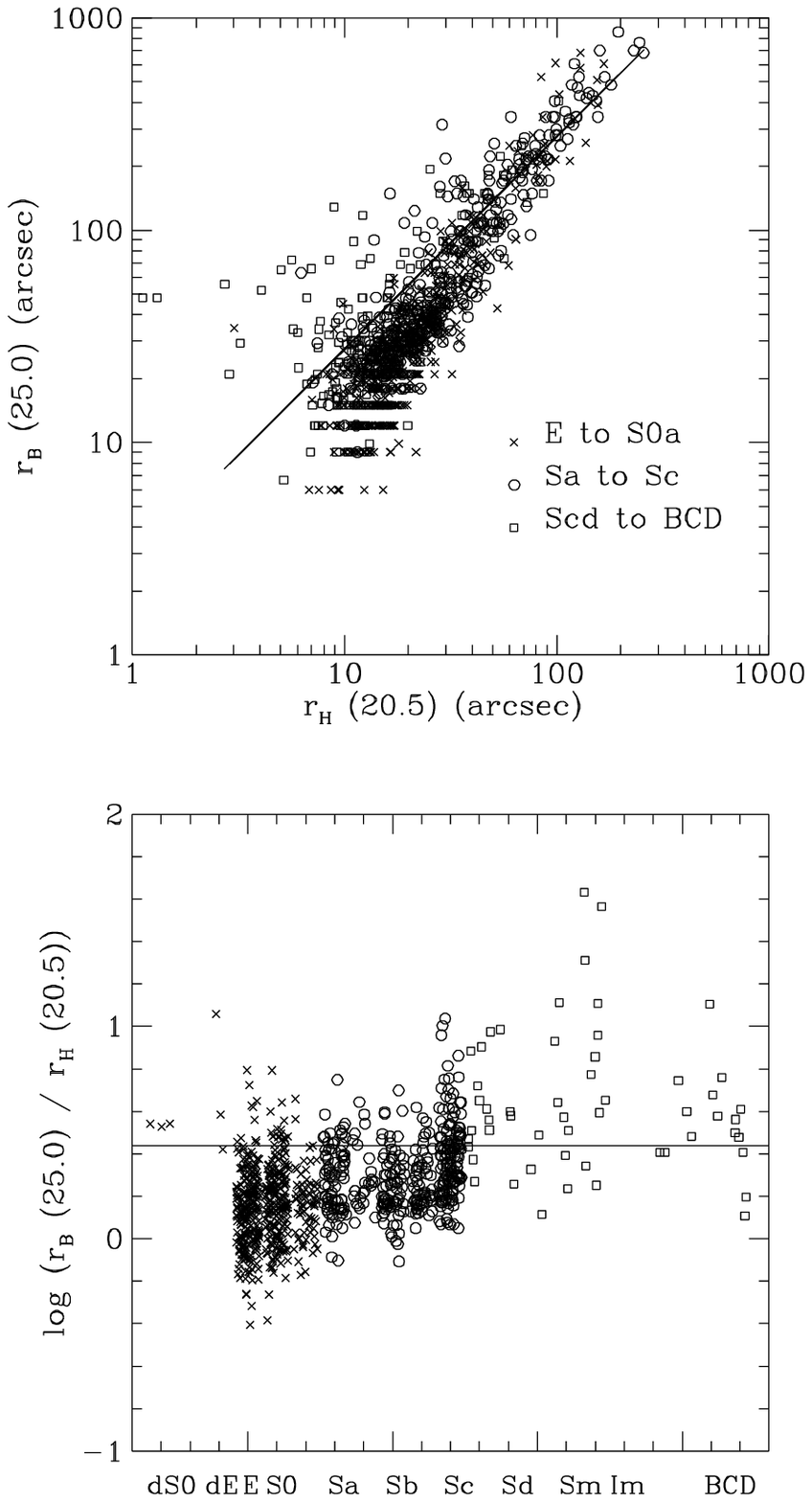,width=15cm,height=15cm}
\caption{The relation between the isophotal optical radius $r_B
(25.0)$ and the infrared one $r_H (20.5)$. The line represents the
linear regression given in the text. Notice that in this and the
following two figures the quantities are plotted in a logarithmic
scale, distorting the distribution of the data-points with respect to
the plotted linear fit.}
\label{fig.10}
\end{figure}

\begin{figure}
\psfig{figure=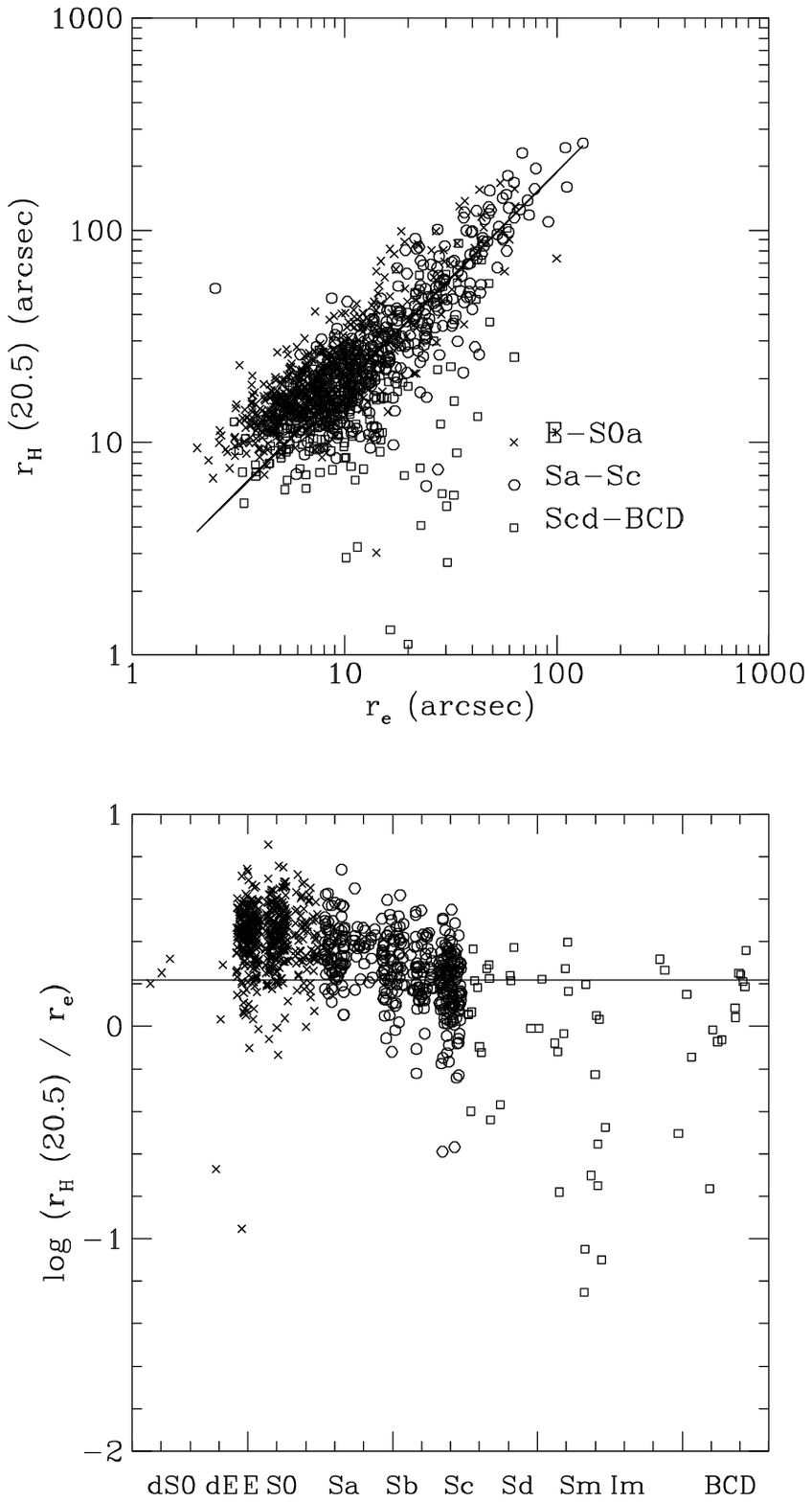,width=15cm,height=15cm}
\caption{The relation between the infrared isophotal radius $r_H
(20.5)$ and the effective radius $r_e$. The line represents the linear
regression given in the text.}
\label{fig.11}
\end{figure}

\begin{figure}
\psfig{figure=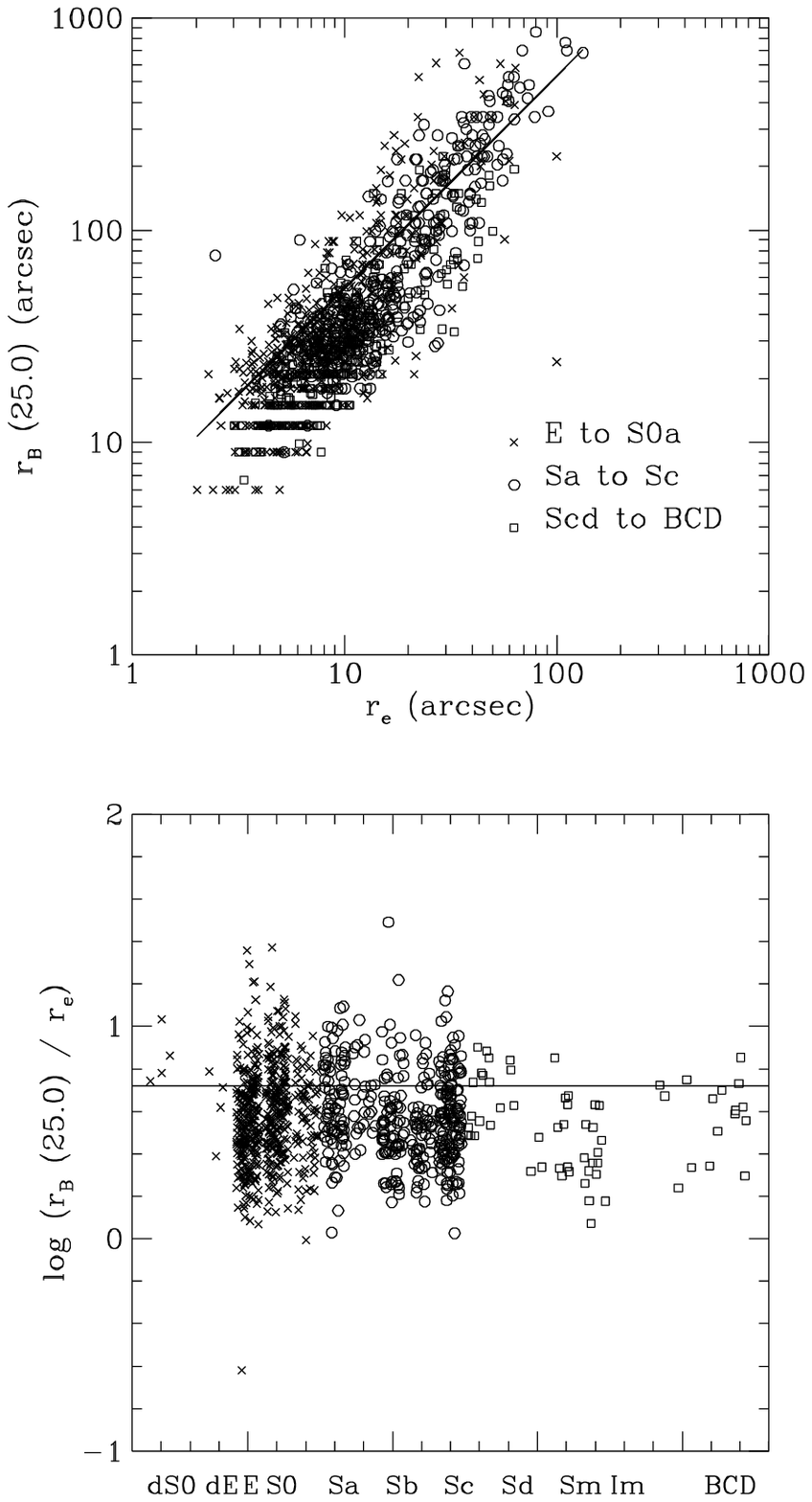,width=15cm,height=15cm}
\caption{The relation between the optical isophotal radius $r_B
(25.0)$ and the effective radius $r_e$. The line represents the linear
regression given in the text.}
\label{fig.12}
\end{figure}

The NIR isophotal radius $r_H (20.5)$ derived at the 20.5 \marc~H-band
isophote, the NIR empirical $r_e$ and the optical isophotal radius
$r_B (25.0)$ are compared in Figs. 10, 11 and 12 (upper panels). The
ratios between these quantities are plotted against the morphological
type in the lower panels.  The morphological type is coded according
to Binggeli et al. (1985).  For plotting purposes we transform the
morphological codes into numeric values and, in order to avoid
superposition of points, we add to it a random number between -0.4 and
0.4. The optical isophotal radii are derived at the
25.0 \marc~B-band isophote except for VCC galaxies, where they are
derived at "the faintest visible isophote". We transform these values
to $r_B (25.0)$ using: $log r_B (25.0) = (log r_{faintest} - 0.25) /
0.8$ as given by Binggeli et al. (1995).

The optical isophotal radii are on the average $2.8\pm 0.03$ of the 
NIR ones (see Fig. 10).  However there are significant deviations from
this simple proportionality, that reflect changes in the mean color of
galaxies with morphological type. As expected, the most extreme case is
that of Scd-Irr-BCD galaxies, that have NIR isophotal radii which are
a small fraction (10 \%) of their isophotal optical radii. In our
sample these galaxies are represented by objects belonging to the 
ISO sub sample of Virgo late-type galaxies, that we observed with the
3.5m Calar Alto telescope (see B97).  In spite of the larger aperture
of the telescope and of the doubled integration time, we could hardly
detect these systems.  Their central NIR surface brightness is often
fainter than 20.0 \marc~ (see Fig. 3c in B97), significantly fainter
than in galaxies of earlier morphological type, as these systems have
hardly any old stellar population at their interior.  The same does
not hold in the optical: they have not as faint central surface
brightness compared with galaxies of earlier types.  Among the 200
E-Sc and the 60 Scd-BCD galaxies in the Virgo cluster the average NIR
effective surface brightness, computed separately in these two
morphological bins, is 17.4 and 19.8 \marc~ respectively, i.e. at NIR
wavelengths the Irr galaxies are 2.4 \marc~ fainter than giant
galaxies of earlier type!  On the contrary, the same galaxies have a
total average optical surface brightness (B magnitude divided by the
optical area) of 22.7 and 23.0 \marc~ respectively, thus showing a
modest 0.3 \marc~ difference.\\ 
At the opposite end of the morphological types sequence are small
early-type galaxies, which have
NIR isophotal radii often exceeding the optical ones.  The latter
evidence is only partly an artifact of the seeing (the deviation from
linearity reduces slightly when the E galaxies observed under the
worse seeing conditions are removed).  Fig. 10 shows that the ratio
$r_B (25)$ / $r_H (20.5)$ increases significantly along the Hubble
sequence, reflecting a genuine color dependence on the Hubble type.

A similar dependence on morphological type is present when
comparing among themselves NIR quantities, like NIR isophotal radius
and NIR effective radius. It appears that $r_H (20.5)$ is on average
$1.9 \pm 0.02$ of $r_e$, as shown in Fig. 11, with the ratio
decreasing along the Hubble sequence. Again, the most significant
deviations are present for Irr galaxies, implying that
20.5 \marc~ is a too high limiting isophote for these galaxies.

Fig. 12 shows that the large deviations among late-type galaxies
vanish if the NIR $r_e$ is compared with the optical $r_B (25)$. We
derive $r_B (25) = 5.3(\pm0.08) \times r_e$ on average, and the ratio
$r_B (25)$ / $r_e$ is constant along the Hubble sequence.

\subsection{The frequency of profile decompositions}

Due to the complete character of the present survey (excluding dEs)
the fraction of the various profile decompositions along the Hubble
sequence can be considered representative of galaxies in the local
Universe.  This is shown in Fig.13 for isolated and cluster
galaxies. Virgo galaxies are kept separate from other clusters to
check if their more reliable morphological classification produces
significant differences.  It is apparent that pure de Vaucouleurs
profiles are present only in 40\% of Es and in 30\% of S0s. Their
contribution drops to zero for later types. The exponential profiles
are absent among early types giant systems up to Sab, but their
frequency dominates (60 \%) in dwarf E+S0s and increases from 40\%
(Sc) to almost 100\% for later types.  Intermediate (B+D)
decompositions dominate from E (50 \%), increasing up to 90\% (Sb),
then they drop to zero for later types. 
This distribution does not
show significant environmental differences among the various clusters
and the isolated sample. Notice however that the above comparison
is restricted to types earlier than Sd, because the isolated sample
does not comprise Irregular galaxies. This is due to the fact
that this sample is
sufficiently far away not to contain low surface brightness Sm and Im 
objects, because it belongs to the Coma supercluster.\\
We notice a more
pronounced dependence on the Hubble type of Virgo galaxies
compared with all other environments.  This is
certainly a consequence of the more reliable morphological
classification available for the Virgo cluster (Binggeli et al. 1985)
than for other more distant objects.  One barely significant
difference is the smaller fraction of pure de Vaucouleurs profiles
among giant E galaxies in Virgo. However this difference is due
to the combination of two effects: the Virgo sample includes
intrinsically fainter galaxies which happen to have a lower
fraction of pure de Vaucouleurs profiles. \\
The dependence of the profile decomposition on luminosity is shown in Figs. 14 and 15,
where the relative fraction of profile decompositions is plotted as a
function of the H band luminosity (log
$L_H/L_\odot$=11.36--0.4$H$+2log$D$ (D in Mpc)).
\begin{figure*}
\psfig{figure=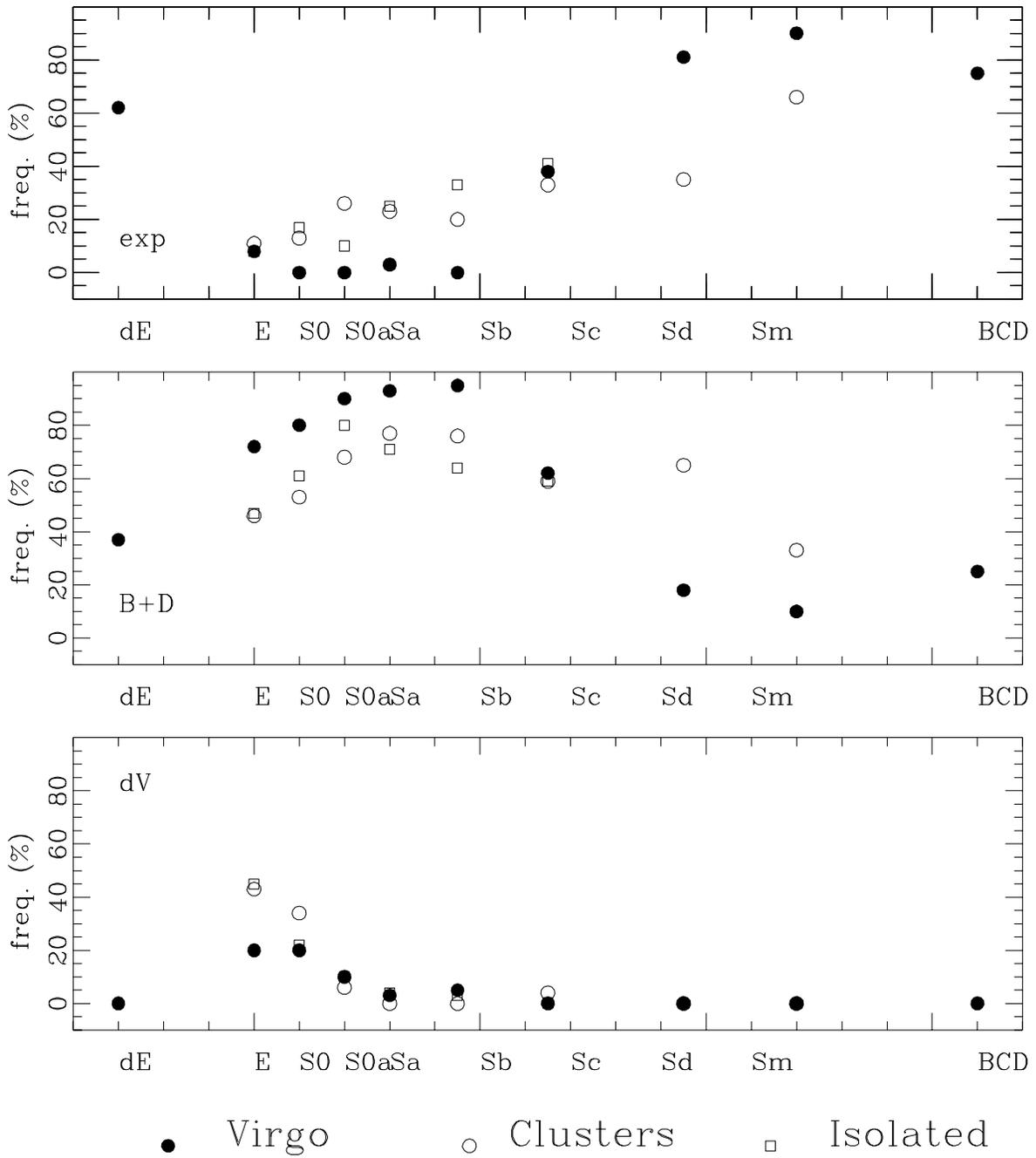,width=19cm,height=19cm}
\caption{The fraction of pure de Vaucouleurs (bottom), pure
exponential (top) and mixed profiles (middle) along the Hubble
sequence is given separately for galaxies in the Virgo cluster, in
other clusters (A262, Cancer, Coma and A1367) and for "isolated"
objects in the Great Wall.}
\label{fig.13}
\end{figure*}
\begin{figure*}
\vskip -8truecm
\psfig{figure=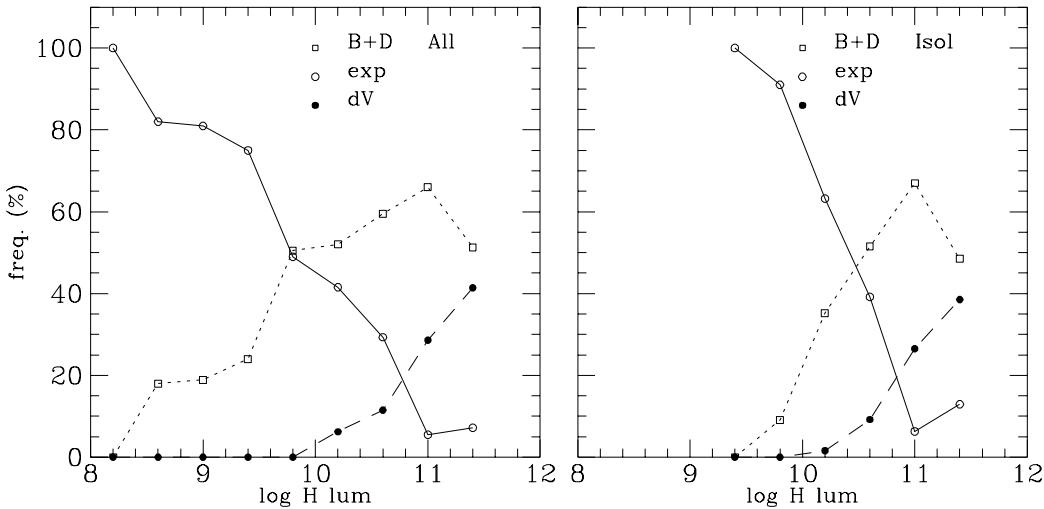,width=17cm,height=17cm}
\caption{The fraction of pure de Vaucouleurs, pure exponential and 
mixed profiles as a function of the NIR luminosity among all galaxies
in the present study (left) and among the "Isolated" objects in the
Coma supercluster (right).}
\label{fig.14}
\end{figure*}
\begin{figure*}
\psfig{figure=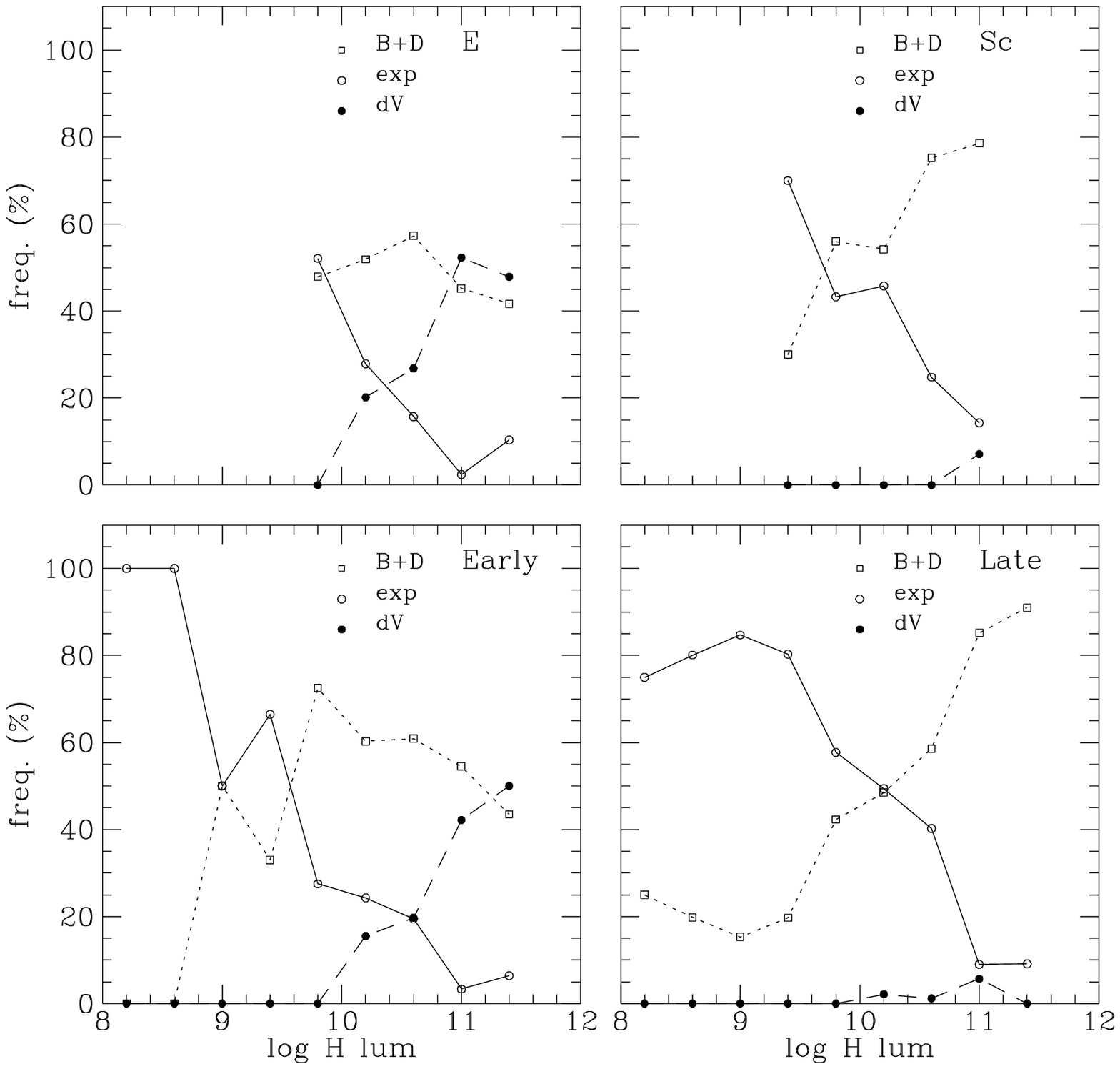,width=17cm,height=17cm}
\caption{The fraction of pure de Vaucouleurs, pure exponential and 
mixed profiles as a function of the NIR luminosity in two broad Hubble type
classes: E+S0+S0a (bottom left panel); Spirals (bottom right) and among
the Ellipticals (top left) and Sc galaxies (top right).}
\label{fig.15}
\end{figure*}
Fig. 14 (left panel) shows that all low luminosity (Dwarf) galaxies have exponential profiles,
while the fraction of B+D decompositions increases with
luminosity. At intermediate luminosities ($L_H=10^{10}$ solar) the two
have an equal frequency (about 50\%).  The pure de Vaucouleurs
profiles are absent below $L_H=10^{10}$ solar and become dominant only
at the highest luminosities.  
This result is independent of the environment, since the same pattern
exists for isolated objects (right panel).\\
The dependence of the profile decomposition on luminosity is basically
morphology-independent (see Fig. 15). In fact a pattern similar to that 
of Fig. 14 is found subdividing the whole sample in two broad
type classes: the E+S0+S0a (Early) versus the Spirals (Late),
and even in two narrow Hubble type classes: Elliptical versus
Sc galaxies.\\
To summarize, we find a dependence of the profile decomposition on
the Hubble type, which is an obvious consequence of the fact that 
the presence or absence of significant bulges enters directly in the
Hubble classification.
Independently from the Hubble classification we find a strong  
correlation between the profile decomposition and the luminosity.
The two relations would not be independent if Hubble type
and luminosity were found correlated one another.
This has been shown not to be the case by Sandage, Binggeli and Tammann
(1985). In their Fig. 21 these authors show that galaxies in the Virgo cluster
have consistent B band luminosity functions in a broad range of morphological
types from giant E to Sc. Only for types later than Sc the average
luminosities are significantly fainter. This feature is present also
in our sample, as shown in Fig. 16. This figure represents the H band
luminosity function of our entire sample, in bins of equal Hubble type.

\section{Summary}

We obtained near-infrared H-band (1.65\micron ) profile decompositions
of 1157 galaxies in five nearby clusters of galaxies: Coma, A1367,
Virgo, A262 and Cancer and in the bridge between Coma and A1367 in the
"Great Wall", taken as representative of isolated galaxies.  
The optically selected ($m_p\leq 16.0$) sample is
representative of all Hubble types, from E to Irr+BCD, except dE.\\ 
We model the surface brightness profiles with the de Vaucouleurs
$r^{1/4}$ law (dV), or with the exponential law (E) or with a
combination of the two (B+D).  Using the fitted quantities we find
that: \\ 
1) The H band effective radii are on average 0.3 of the B band radii
(as determined at the $25^{th}$ \marc B isophote):
$<r_e>=0.18(\pm0.08) \times r_{B25}$.  The ratio $r_B (25)$ / $r_e$ is
constant along the Hubble sequence.\\ 
2) Less than 50\% of the Elliptical galaxies have pure dV profiles.
This is in agreement with the I band surface brightness study by
Scodeggio, Giovanelli \& Haynes (1998).  \\ 
The majority of E to Sb galaxies is best represented by a B+D profile.
Scd-BCD galaxies have pure exponential profiles. \\ 
3) The type of decomposition is a strong function of the total H band
luminosity ($10^8<L_H<10^{11.5}$ \lsol), independent of the Hubble
classification: the fraction of pure exponential decompositions
decreases with increasing luminosity, that of B+D increases with
luminosity.  Pure dV profiles are absent in the low luminosity range
$L_H<10^{10}$ \lsol and become dominant above $10^{11}$ \lsol.\\
\begin{figure}
\psfig{figure=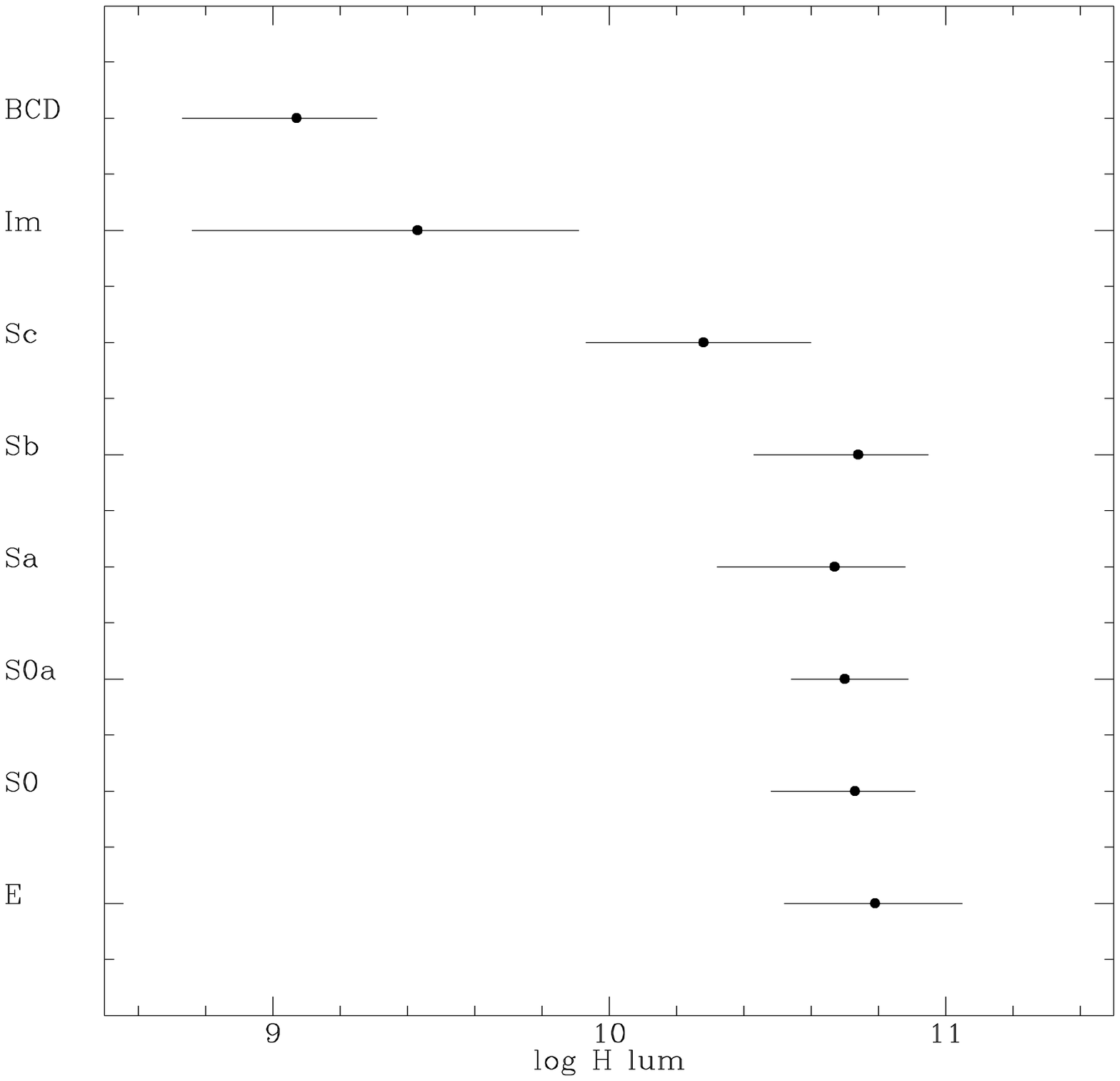,width=10cm,height=10cm}
\caption{The H band luminosity function (HLF) in bins of Hubble type.
Each line represents the extent of the HLF from the 25 to 75 percentile, and the dot
marks the 50 percentile value.}
\label{fig.16}
\end{figure}

\acknowledgements

We wish to thank the TIRGO and Calar Alto T.A.C. for the generous time 
allocation to this project, and Martha Haynes, John Salzer, and
Wolfram Freudling for the development of the GALPHOT package.

\end{document}